\documentstyle[12pt,epsfig]{article}

\catcode`\@=11
\def\@citex[#1]#2{%
\if@filesw \immediate \write \@auxout {\string \citation {#2}}\fi
\@tempcntb\m@ne \let\@h@ld\relax \def\@citea{}%
\@cite{%
  \@for \@citeb:=#2\do {%
    \@ifundefined {b@\@citeb}%
      {\@h@ld\@citea\@tempcntb\m@ne{\bf ?}%
      \@warning {Citation `\@citeb ' on page \thepage \space undefined}}%
      {\@tempcnta\@tempcntb \advance\@tempcnta\@ne%
      \@tempcntb\number\csname b@\@citeb \endcsname \relax%
      \ifnum\@tempcnta=\@tempcntb 
        \ifx\@h@ld\relax%
          \edef \@h@ld{\@citea\csname b@\@citeb\endcsname}%
        \else%
          \edef\@h@ld{\ifmmode{-}\else--\fi\csname b@\@citeb\endcsname}%
        \fi%
      \else
        \@h@ld\@citea\csname b@\@citeb \endcsname%
        \let\@h@ld\relax%
      \fi}%
    \def\@citea{,\penalty\@highpenalty\,}%
  }\@h@ld
}{#1}}

\def\@citeb#1#2{{[#1]\if@tempswa , #2\fi}}
\def\@citeu#1#2{{$^{#1}$\if@tempswa , #2\fi }}
\def\@citep#1#2{{#1\if@tempswa , #2\fi}}

\def\bcites{         
        \catcode`\@=11
        \let\@cite=\@citeb
        \catcode`\@=12
}

\def\upcites{         
        \catcode`\@=11
        \let\@cite=\@citeu
        \catcode`\@=12
}

\def\plaincites{      
        \catcode`\@=11
        \let\@cite=\@citep
        \catcode`\@=12
}

\newcount\hour
\newcount\minute
\newtoks\amorpm
\hour=\time\divide\hour by 60
\minute=\time{\multiply\hour by 60 \global\advance\minute by-\hour}
\edef\standardtime{{\ifnum\hour<12 \global\amorpm={am}%
        \else\global\amorpm={pm}\advance\hour by-12 \fi
        \ifnum\hour=0 \hour=12 \fi
        \number\hour:\ifnum\minute<10 0\fi\number\minute\the\amorpm}}
\edef\militarytime{\number\hour:\ifnum\minute<10 0\fi\number\minute}

\def\draftlabel#1{{\@bsphack\if@filesw {\let\thepage\relax
   \xdef\@gtempa{\write\@auxout{\string
      \newlabel{#1}{{\@currentlabel}{\thepage}}}}}\@gtempa
   \if@nobreak \ifvmode\nobreak\fi\fi\fi\@esphack}
        \gdef\@eqnlabel{#1}}
\def\@eqnlabel{}
\def\@vacuum{}
\def\marginnote#1{}
\def\draftmarginnote#1{\marginpar{\raggedright\scriptsize\tt#1}}
\def\draft{
        \pagestyle{plain}
        \overfullrule=2pt
        \oddsidemargin -.5truein
        \def\@oddhead{\sl \phantom{\today\quad\militarytime} \hfil
        \smash{\Large\sl DRAFT} \hfil \today\quad\militarytime}
        \let\@evenhead\@oddhead
        \let\label=\draftlabel
        \let\marginnote=\draftmarginnote
        \def\ps@empty{\let\@mkboth\@gobbletwo
        \def\@oddfoot{\hfil \smash{\Large\sl DRAFT} \hfil}
        \let\@evenfoot\@oddhead}
        \def\@eqnnum{(\theequation)\rlap{\kern\marginparsep\tt\@eqnlabel}%
        \global\let\@eqnlabel\@vacuum}  }

\def\nblack{            
        \def\ZZ{{Z \n{10} Z}}
        \def\NN{{N \n{14} N}}
        \def\CC{{C \n{11} C}}
        \def\RR{{R \n{11} R}}
        \def\QQ{{Q \n{12} Q}}
        \def\PP{{P \n{11} P}}
}

\def\eqalign#1{\null\,\vcenter{\openup\jot\m@th
  \ialign{\strut\hfil$\displaystyle{##}$&$\displaystyle{{}##}$\hfil
      \crcr#1\crcr}}\,}
\def\eqalignno#1{\displ@y \tabskip\centering
  \halign to\displaywidth{\hfil$\@lign\displaystyle{##}$\tabskip\z@skip
    &$\@lign\displaystyle{{}##}$\hfil\tabskip\centering
    &\llap{$\@lign##$}\tabskip\z@skip\crcr
    #1\crcr}}

\def\section{\@startsection {section}{1}{\z@}{3.ex plus 1ex minus
 .2ex}{2.ex plus .2ex}{\large\bf}}
\def\subsection{\@startsection{subsection}{2}{\z@}{2.75ex plus 1ex minus
 .2ex}{1.5ex plus .2ex}{\bf}}        

\def\appendix{{\newpage\section*{Appendix}}\let\appendix\section%
        {\setcounter{section}{0}
        \gdef\thesection{\Alph{section}}}\section}
\def\thefootnote{\arabic{footnote}}
\def\abstract{\if@twocolumn
\section*{Abstract}
\else 
\begin{center}
{\bf Abstract\vspace{-.5em}\vspace{0pt}}
\end{center}
\quotation
\fi}

\catcode`\@=12

\catcode`\@=11
\def\theequation{\arabic{equation}}
\catcode`\@=12

\nblack
\bcites

\nblack

\catcode`\@=11
\def\theequation{\thesection.\arabic{equation}}
\@addtoreset{equation}{section}
\@addtoreset{footnote}{section}
\@addtoreset{footnote}{subsection}
\catcode`\@=12

\newcommand{\beq}{\begin{equation}}
\newcommand{\beqa}{\begin{eqnarray}}
\newcommand{\bega}{\begin{array}}
\newcommand{\ea}{\end{array}}

\newcommand{\eeq}{\end{equation}}
\newcommand{\eeqa}{\end{eqnarray}}
\newcommand{\p}{\partial}

\newcommand{\lra}{\leftrightarrow}
\newcommand{\at}{{\tilde a}}
\newcommand{\M}{{\cal M}}
\newcommand{\Or}{{\cal O}}

\newcommand{\gss}{\gamma_{\sigma \sigma}}
\newcommand{\gst}{\gamma_{\sigma \tau}}
\newcommand{\intl}{\int_0^l d\sigma \,}
\newcommand{\Yd}{{\dot Y^-}}
\newcommand{\Yp}{{(Y^-)'}}

\begin{document}

\begin{titlepage}

\begin{center}
\hfill EFI-02-76\\

\vskip 2.5 cm
{\large \bf Penrose limit and string quantization in $AdS_3 \times S^3$}
\vskip 1 cm 
\renewcommand{\thefootnote}{\fnsymbol{footnote}}
{
Andrei Parnachev%
\footnote{
        andrei@theory.uchicago.edu
        }
and David A. Sahakyan%
\footnote{
        sahakian@theory.uchicago.edu
        }
}
\setcounter{footnote}{0}
\\
\vskip 0.5cm
{\sl Department of Physics and Enrico Fermi Institute,\\
University of Chicago, Chicago, IL 60637, USA\\ }

\end{center}

\vskip 0.5 cm
\begin{abstract}
We consider corrections to the Penrose limit of $AdS_3 \times S^3$ with
NS-NS flux which are due to the terms next to leading order in
inverse radius expansion.
The worldsheet theory of a lightcone string is interacting
due to the presence of quartic terms in the action.
Perturbative corrections to the spectrum are shown to agree 
with the results from the exact quantization in $AdS_3 \times S^3$.

\end{abstract}

\end{titlepage}

\section{Introduction}
String theory in the pp wave backgrounds has received 
a lot of attention recently (see e.g. \cite{BMN,Me,MT,RT}).
These backgrounds are among the few ones with  non-zero
RR fields where string theory is exactly solvable.
pp waves can also be obtained as the Penrose limit of
$AdS_p \times S^q$ spaces, which have holographic descriptions.
In a recent paper \cite{BMN} Berenstein, Maldacena and Nastase (BMN) proposed 
that the $AdS/CFT$ correspondence \cite{M,GKP,W} which relates
string theory in $AdS_5 \times S^5$ to four dimensional
${\cal N}=4$ Super Yang-Mills,  can be used to
obtain the holographic dual of string theory in the pp wave background.
The important result of their paper is the construction of
boundary operators that correspond to excited string states in the bulk.
In order to have finite lightcone momentum in the pp 
wave background, these states must have spacetime energy and spin which 
go to infinity, as the Penrose limit is taken.  
Corresponding operators are shown to have perturbative (in Yang-Mills coupling)
corrections to the lightcone energy which sum up to reproduce exactly the 
spectrum of a string propagating in the pp wave background.
It is interesting to study analogs of BMN construction
for other backgrounds which have holographic duals.
Among such theories, the near horizon limit of D1/D5
system \cite{CM,HMS} takes a special place.
By using S-duality this system can be mapped  \cite{MSt} to the
near horizon limit of $k$ NS5 branes and $p$ fundamental strings \cite{CHSt,DGHR,Ts}.
The resulting closed string theory on  $AdS_3 \times S^3 \times T^4$ 
is exactly solvable when the string coupling is small,
which happens when $p>>k$.
It has been shown \cite{GKSg,KSg} that the boundary superconformal
algebra can be constructed from the worldsheet point of view.
However the details of the boundary theory are not completely understood.
(See e.g. \cite{BORT,GK,AGS,SW1} for some comments on this issue).

Although one is yet to achieve better understanding of D1/D5
boundary theory which may be needed for the construction 
of an analog of BMN correspondence, it is still possible to
relate the theory in the pp wave background, which arises as the Penrose
limit of $AdS_3 \times S^3 \times T^4$, to the exactly
solvable theory in the original space.
It therefore seems natural to take this example  as a playground
for studying various aspects of the Penrose limit in string theory. 
In this paper we simplify the model even further
and consider the bosonic string on $AdS_3 \times S^3 \times \M$
where $\M$ is a unitary CFT of central charge 20, which is
needed to make the string theory critical.
We exhibit physical states in this background which flow
into string states in the corresponding pp wave background as the Penrose limit
is taken.
An analogous correspondence in $AdS_5 \times S^5$ has been 
investigated recently from the sigma model point of view
in Refs. \cite{Gubser:2002tv,Frolov:2002av}.
We also consider next to leading order corrections to
the pp wave background and write down the interacting
worldsheet theory of a lightcone string.
Perturbative corrections to the lightcone string spectrum
are shown to precisely match the results from the string quantization
in $AdS_3 \times S^3 \times \M$.

This paper is organized as follows.
In the next section we study the lightcone string in the
Penrose limit of $AdS_3 \times S^3 \times \M$ in the leading order
in $1/k$ expansion, reproducing the results of \cite{BMN,RT,KP}.
In section 3 we consider next to leading order corrections to the spectrum.
In Section 4 we show that the spectrum is reproduced from
the exact quantization of a string in $AdS_3 \times S^3 \times \M$.
In Section 5 we discuss our results and mention some directions
for future work.

\section{String in the Penrose limit of $AdS_3 \times S^3$
         in the leading order}
Let us briefly review bosonic string theory on $AdS_3 \times S^3 \times \M$
with the NS-NS flux.  (For a more complete discussion
see e.g. \cite{GKSg,KSg}.)
$\M$ here denotes a unitary CFT with the central charge 20.
The simplest example is just $T^{20}$.
In the rest of the paper we will consider $\M$ to be in its
ground state, although the incorporation of excitations is
straightforward.
The metric on  $AdS_3 \times S^3$ is given by
\beq
\label{adsmetric}
  ds^2=k \left(-\cosh^2 \rho dt^2 +d \rho^2+\sinh^2 \rho d \phi^2
         +\cos^2 \theta d \psi^2 + d\theta^2 + \sin^2 \theta d \chi^2 \right),
\eeq
Here and in the rest of the paper we set $\alpha'=1$.
The NS-NS flux is proportional to the volume form
\beq
H=-2 k \sinh 2 \rho \, dt \wedge d \rho \wedge d \phi
    -2 k \sin 2 \theta \, d \psi \wedge d \theta \wedge d\chi.
\eeq
String theory in this background is described by a 
product of bosonic $SL(2)$ and $SU(2)$ WZNW models at levels 
$k_{SL}=k+2$ and $k_{SU}=k-2$ respectively.
The $SL(2)_L \times SL(2)_R$ and $SU(2)_L \times SU(2)_R$ Kac-Moody algebras
are generated by the holomorphic and antiholomorphic currents.
The mode expansion of holomorphic currents is given by
\beqa
J^{a,SL}(z)&=&\sum_{n=-\infty}^\infty {J^{a,SL}_n\over z^{n+1}},
 \qquad a=\pm,3, \\
J^{a,SU}(z)&=&\sum_{n=-\infty}^\infty {J^{a,SU}_n\over z^{n+1}}, 
\qquad a=\pm,3,
\eeqa
and the expression for antiholomorphic currents is similar.
The central charge of the full theory is
\beq
  c= {3  k_{SL} \over   k_{SL}-2}+{3  k_{SU} \over   k_{SU}+2}+20=26.
\eeq
The states fall into representations of the algebra
with the given SL(2) and SU(2) spins $h$ and $j$ and 
$J^{3,SL}_0,{\tilde J}^{3,SL}_0,J^{3,SU}_0,{\tilde J}^{3,SU}_0$ quantum numbers 
$m_{SL},{\tilde m}_{SL},m_{SU},{\tilde m}_{SU}$.
The BRST invariant states must satisfy physical state
conditions which will be studied in Section 4.

The Penrose limit  \cite{Penrose,BFHP} of (\ref{adsmetric}) corresponds 
to the metric that is seen by a  highly boosted particle.
It is defined by rescaling the coordinates
\beqa
   X^+&=&\frac{(t+\psi)}{2}, \qquad X^-=k \frac{(t-\psi)}{2}, \\
   \rho&=&{r \over \sqrt{k}}, \qquad \theta={y \over \sqrt{k}},
\eeqa
and taking the limit $k \rightarrow \infty$.
This transforms  $AdS_3 \times S^3$ metric into
\beqa
ds^2_0&=&-4 dX^+dX^- + ds^2_{AdS,0}+ds^2_{S,0}, \\
ds^2_{AdS,0}&=& -r^2 (dX^+)^2 + dr^2+r^2 d\phi^2, \\
\label{smetric0}
ds^2_{S,0}&=&-y^2 (dX^+)^2+dy^2 +y^2d\chi^2= -|Z|^2 (dX^+)^2+|dZ|^2,
\eeqa
where $ds^2_{AdS,0}$ and $ds^2_{S,0}$ depend on $AdS_3$ and $S^3$ coordinates 
respectively.
In eq. (\ref{smetric0}) we introduced  a complex coordinate $Z=ye^{i\chi}$ for the 
$S^3$ part of the metric.
The $H$ field in the leading order is given by
\beqa
H_0&=&H_{AdS,0}+H_{S,0}, \\
H_{AdS,0}&=&-4 r \, dX^+ \wedge dr \wedge d\phi, \\
H_{S,0}&=&-4 y \, dX^+ \wedge dy \wedge d\chi=-2 i  \, dX^+ \wedge dZ \wedge d \bar Z.
\eeqa
Note that in the leading order the $AdS_3$ metric and
NS-NS flux are the same as their $S^3$ counterparts.
Therefore it is sufficient to consider the $S^3$ part.
The incorporation of $\M$ is not difficult, as it does not
scale under the Penrose limit.

Let us start by doing the lightcone quantization 
in the leading order in $1/k$ expansion.
The background to this order is described by the
metric (\ref{smetric0}) and the $B$ field
\beq
\label{bfield0}
 B=-i dX^+ \wedge ({ \bar Z}  d Z -Z d {\bar Z}).
\eeq
We will be following closely the treatment of \cite{P1}.
Let us denote the worldsheet coordinates by $\tau \in (-\infty, \infty)$
and $\sigma \in [0,l]$.
The lightcone gauge is specified by setting
\beqa
  X^+&=&\tau, \\
  \p_\sigma \gss&=&0, \\
  {\rm det} \gamma_{\alpha \beta}&=&-1, 
\eeqa
where 
\beq
  \gamma^{\alpha \beta}=\left(  \begin{array}{cc}
   -\gss(\tau) & \gst(\tau, \sigma)  \\
   \gst(\tau, \sigma) &  \gss^{-1}(\tau) (1-\gst^2(\tau, \sigma))
   \end{array} \right) 
\eeq
is the worldsheet metric.
The Lagrangian takes the form 
\beqa
\label{l0}
L_0&=&-{1 \over 4 \pi} \int_0^l d \sigma \bigg[ 
   \gss \left( 4 {\dot x}^- {-} |{\dot Z}|^2 {+}|Z|^2 \right)
  -2 \gst \left( 2  (Y^-)' {-} {1 \over 2} \left[
      {\dot Z} {\bar Z}' + {\dot {\bar Z}}  Z' \right] \right) \\ \nonumber
  && \qquad \qquad  +\gss^{-1} (1-\gst^2)  |Z'|^2 
    -i ({\bar Z}  Z'- Z  {\bar Z}') \bigg], 
\eeqa
where dots and primes denote $\p_\tau$ and $\p_\sigma$ respectively,
and $x^-$ denotes the zero mode of $X^-$
\beq
x^-(\tau)={1 \over l} \int_0^l d\sigma X^-(\tau,\sigma),
\eeq
and 
\beq
 Y^-=X^- - x^-.
\eeq
As explained in \cite{P1}, the equation of motion obtained by the
variation of the action with respect to $Y^-$, together with
the leftover gauge freedom $\sigma \rightarrow \sigma+f(\tau)$,
allows one to set $\gst=0$.
One can also treat $\gst$ as a Lagrange multiplier.
Integrating it out relates $Y^-$ and $Z, \bar Z$
\beq\label{ym}
   2  (Y^-)' = {1 \over 2} \left[
      {\dot Z}  {\bar Z}' + {\dot {\bar Z}}  Z' \right].
\eeq
One must supplement this equation with the constraint
which ensures the  single valuedness of $Y^-$
\beq
\label{svy}
  \intl (Y^-)'=0.
\eeq
The Lagrangian becomes
\beq
L_0=
    \int_0^l  d \sigma {1 \over 4 \pi \gss} \left[-4 \gss^2 {\dot x}^- +
      \gss^2 |{\dot Z}|^2-|Z' + i \gss Z|^2 \right].
\eeq
Since $x^-$ does not explicitly enter the Lagrangian, 
the canonical momentum conjugate to it is conserved and
is related to $\gss$ as
\beq
\label{eomxm0}
 p_-=-{4 \gss l \over 4 \pi}.
\eeq
Let us define
\beq
\label{etap}
 \eta \equiv  -{ p_- \over 2}.
\eeq
It will be convenient to choose the gauge 
\beq
\label{gc}
l=2 \pi \eta=-\pi p_-.
\eeq
In the leading order in $1/k$ expansion $\eta=p^+$
and $\gss=1$, but both relations change when next to leading order
corrections are taken into account.
It is convenient to make a change of variables
\beq
\label{zw}
 Z=e^{-i \sigma}W,
\eeq
which introduces a complex field $W$ satisfying
\beq
  W(\sigma+l)=e^{2 \pi i \eta} W(\sigma).
\eeq
Note that this equation is invariant under $p_- \rightarrow p_- +2$.
However we are going to be interested in the states which 
have $p_- \in [0,2]$.
We will see that states of this type
correspond to unitary representations
of the $SL(2) \times SU(2)$ current algebra, which
are constrained by $1/2 < h < (k+1)/2$.
Other unitary representations, with $h$ out of this bound,
are obtained by the action of the spectral flow \cite{MO}.

The Lagrangian in terms of $W$ becomes that of a free string.
The Hamiltonian then becomes
\beq
\label{lchw}
 H_{lc,0}={1 \over 2} \intl \left[
   (8 \pi) |\Pi_W|^2+{1 \over 2 \pi} |W'|^2 \right],
\eeq
where 
\beq
  \Pi_W={1 \over 4 \pi} {\dot {\bar W}}.
\eeq
The equations of motion for $W,\bar W$ are conventional wave equations,
so the fields can be written as a sum of left and right moving modes
\beqa
W&=& W_L(\tau+\sigma)+W_R(\tau-\sigma), \\
{\bar W}&=& {\bar W}_L(\tau+\sigma)+{\bar W}_R(\tau-\sigma),
\eeqa
which satisfy
\beqa
\dot W_L&=&W'_L,\\
\dot W_R&=&-W'_R,
\eeqa
and have the following oscillator expansion
\beqa
\label{expw}
 W_L&=&{i \over \sqrt{2}} \sum  { \alpha_{n-\eta} \over n-\eta}
         e^{ -{i  (n-\eta) (\tau+\sigma) \over \eta}}, \qquad
 W_R={i \over \sqrt{2}} \sum
      { {\tilde \alpha}_{n+\eta} \over n+\eta}
         e^{ -{i (n+\eta)  (\tau-\sigma) \over \eta}},  \\
 {\bar W}_L&=&{i \over \sqrt{2}}\sum 
  { {\bar \alpha}_{n+\eta} \over n+\eta}
        e^{ -{ i (n+\eta)  (\tau+\sigma) \over \eta }}, \qquad
 {\bar W}_R={i \over \sqrt{2}} \sum
      { {\tilde {\bar \alpha}}_{n-\eta} \over n-\eta}
        e^{-{ i (n-\eta)  (\tau-\sigma) \over \eta }}.
\eeqa
The corresponding expressions for canonical momenta are
\beqa
\Pi_W&=&{1 \over 4  \sqrt{2} \pi \eta}\sum \left[
  { {\bar \alpha}_{n+\eta}}
        e^{ -{ i (n+\eta)  (\tau+\sigma) \over \eta }}+
  { {\tilde {\bar \alpha}}_{n-\eta}}
        e^{-{ i (n-\eta)  (\tau-\sigma) \over \eta }}
                 \right], \\
\Pi_{\bar W}&=&{1 \over 4 \sqrt{2} \pi \eta} \sum \left[ { \alpha_{n-\eta}}
         e^{ -{i  (n-\eta) (\tau+\sigma) \over \eta}} +
  { {\tilde \alpha}_{n+\eta}}
         e^{ -{i (n+\eta)  (\tau-\sigma) \over \eta}} 
                 \right]. 
\eeqa
The canonical commutation relations lead to
the following relation for the oscillator modes\footnote{
    Note that similar commutation relations appear
    for the string in the twisted sector of the orbifold \cite{DFMS}.}
\beq
  \left[\alpha_{n-\eta}, {\bar \alpha}_{-m+\eta}\right]=2 (n-\eta) \delta_{n,m}, 
\eeq
\beq
  [{\tilde {\bar \alpha}}_{n-\eta}, {\tilde \alpha}_{-m+\eta}]=2 (n+\eta) \delta_{n,m}.
\eeq
It will be convenient to introduce canonically
normalized creation and annihilation operators 
\beqa
  a_{n+}^\dagger&=&\frac{\alpha_{-n-\eta}}{ \sqrt{2(n+\eta)}},
      \qquad a_{n+}=\frac{{\bar \alpha}_{n+\eta}}{\sqrt{2(n+\eta)}}, \qquad n>0 \\
  a_{n-}^{\dagger}&=&\frac{{\bar \alpha}_{-n+\eta}}{\sqrt{2(n-\eta)}},
      \qquad a_{n-}=\frac{{\alpha}_{n-\eta}}{\sqrt{2(n-\eta)}}, \qquad n>0 \\
  a_0^\dagger&=&\frac{{\alpha}_{-\eta}}{ \sqrt{2 \eta}}, \qquad
  a_0=\frac{{\bar  \alpha}_{\eta}}{ \sqrt{2 \eta}},
\eeqa
\beqa
  \at_{n-}^\dagger&=&\frac{{ \tilde {\bar \alpha}}_{-n-\eta}}{ \sqrt{2(n+\eta)}},
      \qquad \at_{n-}=
    \frac{{{\tilde {\alpha}}}_{n+\eta}}{\sqrt{2(n+\eta)}}, \qquad n>0 \\
 \at_{n+}^{\dagger}&=&\frac{{\tilde {\alpha}}_{-n+\eta}}{\sqrt{2(n-\eta)}},
      \qquad \at_{n+}=
      \frac{{\tilde{\bar \alpha}}_{n-\eta}}{\sqrt{2(n-\eta)}}, \qquad n>0 \\
  \at_0^\dagger&=&\frac{{\tilde {\bar \alpha}}_{-\eta}}{ \sqrt{2 \eta}}, \qquad
  \at_0=\frac{{\tilde \alpha}_{\eta}}{ \sqrt{2 \eta}}.
\eeqa
Pluses and minuses in subscript denote the transformation
properties of the oscillators under the rotation of complex
plane $Z \rightarrow e^{i \theta} Z$.
Substituting the oscillator expansions 
into (\ref{lchw}) and using the commutation relations one can
obtain the following form for the lightcone Hamiltonian
\beqa
\label{lch}
  H_{lc,0}&=& 2+ N'+{\tilde N}'+{N + {\tilde N}-2 \over \eta}= 
    2+ N'+{\tilde N}'-2 {N + {\tilde N}-2 \over p_-} \\
  N&=& \sum_{n>0} n \left[ a_{n+}^\dagger a_{n+}+a_{n-}^\dagger a_{n-} \right], \\
  {\tilde N}&=& \sum_{n>0} n \left[
                  {\tilde a}_{n+}^\dagger {\tilde a}_{n+}+
                   {\tilde a}_{n-}^\dagger {\tilde a}_{n-}  \right],\\
  N'&=& a_0^\dagger a_0+
         \sum_{n>0} \left[ a_{n+}^\dagger a_{n+}-a_{n-}^\dagger a_{n-} \right],\\
  {\tilde N}'&=&\at_0^\dagger \at_0 + \sum_{n>0} \left[ 
      {\tilde a}_{n-}^\dagger {\tilde a}_{n-} 
                  -{\tilde a}_{n+}^\dagger {\tilde a}_{n+} \right].
\eeqa
Additional factors of 2 arise due to normal ordering of
the oscillators\footnote{In computing normal ordering
constants one should keep in mind that the total central charge
of the matter theory is 26 and $\M$ is in its ground state.}.
The Hamiltonian (\ref{lch}) must be supplemented by
a  usual physical state condition,
\beq
\label{zerop}
 N={\tilde N},
\eeq
which is a consequence of (\ref{svy}).
Note that the Hamiltonian (\ref{lch}) contains a tachyonic state
at level 0, but is nonnegative for states with  $N>0$ and $p_- \in [0,2]$.

\section{Next to leading order corrections to the Penrose limit}
Our next task will be computing $\Or(1/k)$ corrections
to the lightcone spectrum (\ref{lch}).
The corrections to the $S^3$ metric (\ref{smetric0})
and  $B$ field (\ref{bfield0}) are 
\beqa
 ds^2_{S,1}&=&{1 \over k} \left[ 2 y^2 dX^+ dX^- + {y^4 \over 3} (d X^+)^2-
         {y^4 \over 3} (d \chi)^2 \right]=\\ \nonumber
  && \qquad   {1 \over k} \left[ 2 |Z|^2 dX^+ dX^- {+} 
    {|Z|^4 \over 3} (d X^+)^2 + {1 \over 12} (Z d{\bar Z}-{\bar Z} dZ)^2 \right],  \\
 B_1&=& {1 \over k} \left[i {|Z^2| \over 3 } dX^+ \wedge ({ \bar Z}  d Z -Z d {\bar Z}) 
    +i dX^- \wedge ({ \bar Z}  d Z -Z d {\bar Z}) \right].
\eeqa
Note that $AdS_3$ metric and $B$ field can be obtained
from their $S^3$ counterparts by the substitution  $k \rightarrow -k$.
Therefore in the calculations below we will only retain
the $S^3$ part.

The correction to the Lagrangian therefore becomes
\beqa
\label{l1}
L_1&=&{-}{1 \over 4 \pi k} \intl \bigg[ 
    {-}2 |Z|^2 {\dot X}^- {-} {1 \over 3} |Z|^4 {-}
    {1 \over 12} (Z {\dot {\bar Z}}{-}{\bar Z} {\dot Z})^2            
    {+}{1   \over 12} (Z {\bar Z}'{-}{\bar Z}  Z')^2 \\
 \nonumber && \qquad \qquad
 +i {|Z|^2 \over 3} \left( {\bar Z}  Z'-Z {\bar Z}' \right)
 +i {\dot X}^- \left( {\bar Z}  Z'-Z {\bar Z}' \right)
 -i  (X^-)' \left( {\bar Z} {\dot Z}-Z {\dot {\bar Z}} \right)
\bigg].
\eeqa
Note that in the leading order in $1/k$
$\gst$ is set to zero by the variation of $Y^-$.
On the other hand, varying $\gst$ implies that
\beq
 \left( 2  (Y^-)' {-} {1 \over 2} \left[
      {\dot Z} {\bar Z}' + {\dot {\bar Z}}  Z' \right] \right) \sim \Or(1/k).
\eeq
Therefore we do not write terms containing $\gst$, as they are
higher order in $1/k$.

It will be convenient to separate the Lagrangian (\ref{l1})
into three terms
\beqa
  L_1&{=}&L_1^{(1)}+L_1^{(2)}+L_1^{(3)}, \\
  L_1^{(1)}&{=}&-{1 \over 4 \pi k} \intl \bigg[ -2  |Z|^2 {\dot x}^- 
   +i {\dot x}^- \left( {\bar Z}  Z'-Z {\bar Z}' \right) \bigg], \\
  L_1^{(2)}&{=}&{-}{1 \over 4 \pi k}  \intl \bigg[ 
     {-} {1 \over 3} |Z|^4 {-}
         {1 \over 12} (Z {\dot {\bar Z}}{-}{\bar Z} {\dot Z})^2 
 \\ \nonumber && \qquad \qquad
  {+}{1 \over 12}( {\bar Z}  Z'{-}Z {\bar Z}' )^2  
 +i {|Z|^2 \over 3} \left( {\bar Z}  Z'-Z {\bar Z}' \right) \bigg],\\
   L_1^{(3)}&{=}& {-}{1 \over 4 \pi k}  \intl \bigg[   {-}2 |Z|^2 {\dot Y}^-
              {+}i {\dot Y}^- \left( {\bar Z}  Z'{-}Z {\bar Z}' \right)
       {-}i  (Y^-)' \left( {\bar Z} {\dot Z}{-}Z {\dot {\bar Z}} \right)
\bigg].
\eeqa
Let us analyze this expression, term by term.
We first consider the effect of the addition of
\beq
  L_1^{(1)}{=}{1 \over 2 \pi k} \intl
    {\dot x}^- \; Im ( {\bar W}  W')
\eeq
to the leading order Lagrangian.
The eq. (\ref{eomxm0}) is modified to
\beq
\label{etapnew}
 p_-=-2 \eta =-{4 \gss l \over 4 \pi} +
       {1 \over 2 \pi k} \intl Im ( 
         {\bar W}_L W'_L +  {\bar W}_R W'_R
        + {\bar W}_L W'_R +  {\bar W}_R W'_L).
\eeq
With the gauge choice (\ref{gc}) this implies
\beqa
\gss&=&1+\delta \gss, \\
\label{dgss}
 \delta \gss &=&  {N'+{\tilde N}' \over 2 k \eta}+
   {1 \over 4 \pi k \eta} \intl \; Im ({\bar W}_L W'_R +  {\bar W}_R W'_L),
\eeqa
where we used the oscillator expansion of fields
to compute the integral 
\beq
\label{usefuli}
\int_0^l {d\sigma \over 2 \pi} \,
          Im (   {\bar W}_L W'_L + {\bar W}_R W'_R )=
N'+{\tilde N}'. 
\eeq
In writing this result we omitted a constant which arises
due to normal ordering of the zero modes.
However when the  $AdS^3$ part is added 
its zero modes come with a relative minus sign and therefore
in the full $AdS_3 \times S^3$ expression the normal ordering
constants will cancel.

At this point we would like to make some remarks 
on how perturbation theory works.
Suppose one adds a small perturbation to the harmonic oscillator Lagrangian
\beq
  L={{\dot x}^2 \over 2}-{w^2 x^2 \over 2}+\alpha f,
\eeq
where $f=f(x,{\dot x})$ and $\alpha$ is a small parameter.
The canonical momentum becomes
\beq
p={\dot x}+\alpha {\p f \over \p {\dot x}}.
\eeq
We can write the new Hamiltonian omitting terms $\Or(\alpha^2)$.
\beq
 H={p^2 \over 2}+{w^2 x^2 \over 2}-\alpha f.
\eeq
Hence to compute the change in the Hamiltonian up to
next to leading order due to $\Or(1/k)$ corrections to
the Lagrangian we just need to subtract these terms from
the unperturbed Hamiltonian.
Since the perturbation is already $\Or(1/k)$, we can 
use the original oscillator expansions for the fields.
The standard quantum mechanical perturbation theory then
implies that the correction to the lightcone energy 
of a given string state is given by the diagonal 
matrix element. 
As explained above, the effect of $L_1^{(1)}$ is to
change $\gss$ by a term $\Or(1/k)$.
This induces the $\Or(1/k)$ correction to $L_0$, which
we denote by $\delta L_0$.
The corresponding correction to the lightcone energy 
is given by
\beqa
\label{lcec1}
H_{lc,1}^{(1)}=-\langle \delta L_0 \rangle&=&\bigg\langle { \delta \gss \over 4 \pi} \intl \left[
     -{\dot {\bar W}} {\dot W}-{\bar W}' W'+2 Im( {\bar W} W') \right] \bigg\rangle=
 \\ \nonumber && 
 \bigg\langle \delta \gss  \left( -{ (N+{\tilde N}-2) \over \eta}+
         \int_0^l {d \sigma \over 2 \pi}
       \; Im ({\bar W}_L W'_R +  {\bar W}_R W'_L) \right) \bigg\rangle.
\eeqa
In obtaining this formula we expressed $\delta L_0$ in terms of
$W,\bar W$ and used the oscillator expansions.
Substituting (\ref{dgss}) and using the fact that the
expectation value is taken with respect to the eigenstate
of oscillator  number operators, we can rewrite (\ref{lcec1})
as 
\beq
\label{h1}
   H_{lc,1}^{(1)}=-{ (N+{\tilde N}-2) (N'+{\tilde N}') \over 2 k \eta^2}
    +{1 \over 2 \eta k} \bigg\langle \left( \int_0^l {d \sigma \over \pi}
       \; Im [W_L {\bar W}'_R]  \right)^2 \bigg\rangle,
\eeq
where we used
\beq
\intl Im(W_L\bar W'_R{-}W'_L \bar W_R)=\intl (Im(2W_L\bar W'_R{-}(W_L \bar W_R)')=
\intl Im(2W_L\bar W'_R).
\eeq
It is easy to compute the contribution of $L^{(2)}$ to the lightcone energy
\beq
 H_{lc,1}^{(2)}=-\langle L^{(2)} \rangle=\bigg\langle
      {1 \over 4 \pi k} \intl {1 \over 12} \left[
      ({\bar W} W'-W {\bar W}')^2- ({\bar W} {\dot W}-W {\dot {\bar W}})^2 \right]
     \bigg\rangle.
\eeq
Substituting the explicit form of $L^{(2)}$, 
using the free field solution for $W, {\bar W}$ and
taking advantage of the fact that we need to compute 
an expectation value, we have
\beq
\label{h2}
 H_{lc,1}^{(2)}=\bigg\langle {1 \over 2 \pi k} \intl W_L {\bar W_L}' W_R {\bar W}'_R \bigg\rangle.
\eeq
 

The change in the light cone energy 
due to $L^{(3)}_1$ term in the Lagrangian is
\beqa
H_{lc,1}^{(3)}=-\langle L^{(3)} \rangle&=&\bigg\langle {1\over 2\pi k}\intl\bigg [{-}|W|^2\Yd{+}
Im(\dot W\bar W)\Yp{-}\\
&&\nonumber\qquad \qquad Im(\bar W ({-}iW{+}W') )\Yd\bigg]\bigg\rangle=\\
&&\nonumber  \bigg\langle {1\over 2\pi k}\intl\left[Im(\dot W\bar W)\Yp{-}
Im(W' \bar W )\Yd\right]\bigg\rangle.
\eeqa
In order to compute this correction we shall use (\ref{ym}) to express 
$\Yp$ and $\Yd$ in terms of oscillator modes.
The eq. (\ref{ym})  can be written as
\beq\label{ym1}
2\Yp=Re[\dot W(i\bar W+{\bar W}')].
\eeq
It will be convenient to express $W,\bar W$ in terms of left and right moving modes 
$W_L, \bar W_L$ and $W_R, \bar W_R$. 
Then (\ref{ym1}) becomes
\beq
2\Yp{=}{-}Im(W'_L \bar W_L){+}W'_L\bar W'_L{+}
Im(W'_R \bar W_R){-}W'_R\bar W'_R{-}
 Im(W_L\bar W_R)'.
\eeq
We see that $\Yp$ can be written as
\beq
\Yp=\Yp_L+\Yp_R+\Yp_{LR},
\eeq
where
\beqa
\Yp_L&=&{1\over 2}({-}Im(W'_L \bar W_L){+}W'_L\bar W'_L),\\
\Yp_R&=&{1\over 2}(Im(W'_R \bar W_R){-}W'_R\bar W'_R),\\
\Yp_{LR}&=&-{1\over 2}\p_{\sigma}Im(W_L\bar W_R).
\eeqa
Now one can easily find 
\beq\label{ydot1}
\Yd=\Yp_L-\Yp_R-{1\over 2}\p_{\tau}Im(W_L\bar W_R)+a(\tau),
\eeq
where $a(\tau)$ should be chosen such that
\beq\label{ydot}
\intl\Yd=0,
\eeq
which is a consequence of the definition of $Y^-$,
\beq
\intl Y^-=0.
\eeq
One can write down an explicit formula for $a(\tau)$,
\beqa
a(\tau)&=&-{1\over l}\intl\left(\Yp_L-\Yp_R-
{1\over 2}Im(W'_L\bar W_R-W_L\bar W'_R)\right)=\\
&&\nonumber\qquad
-{(N+\tilde N-2)\over 2\eta^2}-{1\over 2l}\intl Im(W_L\bar W'_R-W'_L \bar W_R).
\eeqa
Now we are ready to compute the change in the light cone energy
(we are omitting some terms whose expectation 
value is vanishing )
\beqa
H_{lc,1}^{(3)}&=&-\bigg\langle {1\over 4\pi k}\intl \bigg[2[W'_L\bar W'_L-Im(W'_L\bar W_L)]
Im(W'_R\bar W_R)+\\ \nonumber \\ \nonumber 
&& \qquad \qquad 2[W'_R\bar W'_R{-}Im(W'_L\bar W_R)]
Im(W'_R\bar W_R){-}4Im(\bar W_R W'_L)Im(\bar W_LW'_R) \bigg] \bigg\rangle -\\
\nonumber\\
&&\nonumber \bigg\langle
\bigg({N+\tilde N-2\over \eta^2}+{1\over l}\intl Im(W_L\bar W'_R-W'_L\bar W_R)\bigg) \times \\ \nonumber \\ \nonumber &&
\qquad\qquad  {1\over 4\pi k} \intl
  Im(W'_L\bar W_L+W'_R\bar W_R+\bar W_LW'_R+\bar W_RW'_L)\bigg\rangle. 
\eeqa
After some algebra one can arrive to the following expression
\beqa\label{h3}
H_{lc,1}^{(3)}=-{1\over k\eta^2}\bigg\langle \left((N-1)\tilde N'+(\tilde N-1)N'-
{1\over 2}(N+\tilde N-2)(N'+\tilde N')\right) \bigg\rangle-\\
\nonumber\bigg\langle {1\over 2 \pi k}
  \intl W'_L \bar W_L W'_R \bar W_R{+}{1\over 2 k \eta} 
    \left(\int_0^l {d \sigma \over \pi} Im(\bar W_L W'_R)\right)^2
\bigg\rangle,
\eeqa
where we used $l=2 \pi \eta$.
The term in the first line vanishes due to condition 
\beq
N=\tilde N.
\eeq
The first term in the second line completely cancels
(\ref{h2}), while the second term in this line cancels
the second term in (\ref{h1}).
Hence the the lightcone spectrum up to next to leading order in $1/k$ is 
\beqa
\label{lchnew}
  H_{lc}&=&H_{lc,0}+\sum_{i=1}^3 \, H_{lc,1}^{(i)}= \\ \nonumber
        && 2+N'+{\tilde N}'-{2( N+{\tilde N}-2) \over p_-} 
          - {2 (N+{\tilde N}-2) (N'+{\tilde N}')\over  k (p_-)^2 }.
\eeqa
This can also be written as
\beq
 H_{lc}= 2+N'+{\tilde N}'-{4(N-1) \over p_-} 
          - { 4 (N-1) (N'+{\tilde N}')\over  k (p_-)^2 }\\
\eeq
due to the physical state condition.
This however is not the end of the story, as one must
remember that the original space contains a compact
direction $\psi$ which implies quantization
of the SU(2) spin.
Corresponding Killing vector is 
\beq
  -i \p_{\psi}={ -i \p_{X^+} - k (-i \p_{X^-}) \over 2}.
\eeq
The momentum which generates translations in $\psi$ has to be
quantized, which implies
\beq
   {p_+-k p_- \over 2} =- {k p_- + H_{lc} \over 2} \in Z.
\eeq
Substituting (\ref{lchnew}) and retaining terms up to next to 
leading order recasts this constraint as
\beq
\label{constraint}
  -{k p_- \over 2} - 1-{N'+{\tilde N}' \over 2}+{2 (N-1) \over p_-} =-i \p_\psi \in Z
\eeq

\section{$AdS_3 \times S^3$: exact quantization}
Let us study states with $j, h \sim \Or(k)$ in the large $k$ limit.
In the following we will keep terms up to the second order
in the expansion in powers of $1/k$.
The physical state condition for states with oscillator
number $N$
\beq
\label{psc}
  -\frac{h (h-1)}{k} +\frac{j (j+1)}{ k}+N=1,
\eeq
relates $h$ and $j$ as
\beq
\label{hj}
h=j+1+{(N-1) k \over 2 j}-
 {1 \over k} { (N-1) k \over 2 j}\left( { (N-1) k \over 2 j}+1\right).
\eeq  
We will generate all states from the primary
state $|j \rangle$ which has the SL(2) spin $h$ related to 
the SU(2) spin $j$ as in (\ref{hj}).
The quantum numbers of this state are given by
\beq
  m^{SL}= {\tilde m}^{SL}=h, \qquad m^{SU}= {\tilde m}^{SU}=j.
\eeq
These quantum numbers are related to $\eta$ and $H_{lc}$,
which appear in the Penrose limit, as follows
\beqa
\label{lcep}
H_{lc}=m^{SL}+{\tilde m}^{SL}-m^{SU}-{\tilde m}^{SU}, \\
\label{lcmp}
2 \eta=-p_-={m^{SL}+{\tilde m}^{SL}+m^{SU}+{\tilde m}^{SU} \over k}.
\eeqa
It was argued in \cite{BMN} that  bulk isometries
correspond to string zero modes in pp wave background.
In the present  context this implies that the action of the
zero modes of the SU(2) current algebra  $J^{-,SU}_0,{\tilde J}^{-,SU}_0$
on the state $|j\rangle$ corresponds to the action of zero mode 
creation operators $a_0^\dagger, {\tilde a}_0^\dagger$ on the lightcone
vacuum in the pp wave background.
In other words, we have the correspondence
\beqa
   a_0^\dagger &\lra&  J^{-,SU}_0, \qquad  a_0 \lra  J^{+,SU}_0, \\
   {\tilde a}_0^\dagger  &\lra&  {\tilde J}^{-,SU}_0, \qquad  
           {\tilde a}_0 \lra  {\tilde J}^{+,SU}_0.
\eeqa
This correspondence extends to the nonzero modes
\beqa
   a_{n+}^\dagger &\lra& J^{-,SU}_{-n}, \qquad  a_{n+} \lra J^{+,SU}_{n},\\
   a_{n-}^\dagger &\lra& J^{+,SU}_{-n}, \qquad  a_{n+} \lra J^{-,SU}_{n}, \\
   {\tilde a}_{n+}^\dagger &\lra& {\tilde J}^{+,SU}_{-n}, \qquad
             {\tilde a}_{n+} \lra {\tilde J}^{-,SU}_{n}, \\
   {\tilde a}_{n-}^\dagger &\lra& {\tilde J}^{-,SU}_{-n}, \qquad 
              {\tilde a}_{n-}^\dagger \lra {\tilde J}^{+,SU}_{n},
\eeqa
with similar correspondence for the $AdS_3$ part.
It is easy to see that this identification implies 
correct matching for the transformation properties
under the rotation in the $Z$ plane generated by $J^{3,SU}_0-{\tilde J}^{3,SU}_0$.
Furthermore, the physical state condition (\ref{zerop}) is
naturally interpreted as level matching on the $AdS_3 \times S^3 \times \M$ 
side.
Now we would like to show that the lightcone energy 
of the physical states precisely reproduces the result
obtained from lightcone quantization (\ref{lchnew}), up to
next to leading order in $1/k$ expansion.
Let us first note that
\beq
  m^{SL}={\tilde m}^{SL}=h, \qquad m^{SU}=j-N', \qquad {\tilde m}^{SU}=j-{\tilde N}'.
\eeq
Using this relation, together with (\ref{lcep}) and (\ref{hj}) we obtain
\beq
\label{lche}
  H_{lc}=2+N'+{\tilde N}'+{2 (N-1) k \over 2 j}-
         {2 (N-1) \over k} \, \left({k \over 2 j} \right)^2 \left(
         {(N-1) k \over 2 j}+1 \right).
\eeq
One can also write down the relation between 
$p_-$ and $j$ using (\ref{lcmp}) and (\ref{hj}):
\beq
  -p_-={1 \over k} \left[2 + 4 j +{2 (N-1) k \over 2 j} -N' -{\tilde N}' \right],
\eeq  
which implies
\beq
\label{jp}
 2 j=-{k p_- \over 2}-1 +{2(N-1) \over p_-} +{N'+ {\tilde N}' \over 2}
\eeq
up to terms which are $\Or(1/k)$.
Note that the requirement for $2 j$ to be an integer 
reproduces precisely the constraint (\ref{constraint}) which
appeared due to the compactness of the space.
In fact, what should appear in the right hand side of
(\ref{constraint}), is 
\beq
 -i \p_ \psi=2 j-(N'+ {\tilde N}'),
\eeq
meaning that (\ref{constraint}) and (\ref{jp}) are completely
equivalent.
Substituting (\ref{jp}) into (\ref{lche}) reproduces the
lightcone Hamiltonian (\ref{lchnew}).

\section{Discussion}
In this paper we studied bosonic string propagating
in the Penrose limit  of $AdS_3 \times S^3 \times \M$.
We identified states in the leading order in $1/k$
expansion (that is, states in the pp wave background) with
physical states in $AdS_3 \times S^3 \times \M$.
When one goes beyond the leading order, the pp wave background
receives corrections.
The worldsheet theory of a lightcone string propagating there becomes interacting, with the
appearance of quartic terms in the action.
We computed next to leading order corrections to the
spectrum in perturbation theory and found precise 
agreement with the results from exact quantization 
in original $AdS_3 \times S^3 \times \M$ background.
The correspondence has been worked out for states in
the unitary representations with $AdS_3$ spin $1/2<h<(k+1)/2$.
The incorporation of spectrally flowed states should be 
relatively simple.
One should also be able to extend the results of this paper
to superstring case. 
Let us mention some other interesting problems and directions
for future work.
We would like to reiterate that $AdS_3 \times S^3$ background,
because of its solvability, seems to be a good toy model for studying 
various features of the Penrose limit in string theory.
For example, one may be able to see directly how correlation
functions behave under the Penrose limit both in the bulk 
and in the boundary theory.
Perhaps one can also make some progress in constructing
a holographic dual to the pp wave background.

One of the most interesting questions raised by BMN construction
is why operators that correspond to the excited string states 
take the form of single trace operators with phase insertions \cite{BMN}.
It is possible that  $AdS_3 \times S^3$ example will be useful
in understanding this better.
However it is likely that pursuing this question will require
generalization of the result of this paper to the case of
nonzero RR flux, in order to be able to study a nonsingular
boundary theory.
In fact, it is not hard to construct operators in the boundary 
theory that correspond to states that are created by the action of
zero mode oscillators on the lightcone vacuum in pp wave background.
The reason is that these states essentially correspond to
chiral primary observables in the boundary SCFT which are
independent of moduli.
Corresponding worldsheet operators have been constructed
for the theory with NS-NS flux in \cite{KLL} and they agree with
the spectrum of chiral primaries in  $(T^4)^{k p}/S_{k p}$ orbifold \cite{LM},
even though the two theories correspond to different points in
the moduli space of D1/D5 system.

At any rate, it seems like generalization of our results
to $AdS_3 \times S^3$ background with RR flux deserves further attention.
In this case the exact quantization scheme is available \cite{Berkovits:1999im}.
Another interesting direction would be studying
other backgrounds with RR flux, where the exact quantization procedure
is lacking.
Note that we worked out first two leading terms in the $1/k$
expansion, but in principle one can go further and extend
the results to higher orders in perturbation theory.
This procedure would provide an alternative way of quantizing
string theory in $AdS_3 \times S^3 \times \M$ and may prove
very useful when applied to backgrounds that cannot be quantized
exactly.
Even in  $AdS_3$ we may learn something new, as the conventional 
quantization procedure deals with the Euclidean version
of $AdS_3$, and the results of analytic continuation are
not completely understood.
For example, one may try to understand better D-branes
in Lorentzian $AdS_3$ and make a connection with 
algebraic results of \cite{GKS,PS,PST,LOP}.
D-branes in pp wave backgrounds have been investigated
recently in \cite{DP,KNS,Bak,ST,Billo:2002ff}.

\section{Acknowledgements}
We would like to thank Emil Martinec, Vasilis Niarchos, Anton Ryzhov
and especially David Kutasov for useful discussions.
This work was supported, in part, by DOE grant \#DE-FG02-90ER40560.



\begin{thebibliography}{99}


\bibitem{BMN} D. Berenstein, J. Maldacena, H. Nastase,
``Strings in flat space and pp waves from N=4 Super
Yang Mills'', hep-th/0202021;

\bibitem{Me} 
R.~R.~Metsaev,
``Type IIB Green-Schwarz superstring in plane wave Ramond-Ramond  background'',
Nucl.\ Phys.\ B625, 70 (2002)
[hep-th/0112044];

\bibitem{MT} R.R. Metsaev, A.A. Tseytlin, ``Exactly solvable model
of superstring in plane wave Ramond-Ramond background'', hep-th/0202109;

\bibitem{RT} J.G. Russo, A.A. Tseylin, ``On solvable
models of type IIB superstrings in NS-NS and R-R sector 
backgrounds'', hep-th/0202179;

\bibitem{M} J. Maldacena, ``The large N limit of superconformal field
theories and supergravity'', Adv. Theor. Math. Phys. 2, 231 (1998)
[hep-th/9711200];

\bibitem{GKP} S.S. Gubser, I.R. Klebanov and A.M. Polyakov, 
``Gauge theory correlators from non-critical string theory'',
Phys. lett. B428, 105 (1998) [hep-th/9711200];

\bibitem{W} E. Witten,  ``Anti-de Sitter space and holography'',
Adv. Theor. Math. Phys. 2, 253 (1998) [hep-th/9802150];

\bibitem{CM} C. Callan, J. Maldacena, 
``D-brane Approach to Black Hole Quantum Mechanics'',
Nucl.Phys. B472 591 (1996) [hep-th/9602043];

\bibitem{HMS} G. Horowitz, J. Maldacena, A. Strominger,
``Nonextremal Black Hole Microstates and U-duality'', Phys.Lett. 
B383, 151 (1996) [hep-th/9603109];

\bibitem{MSt} J. Maldacena, A. Strominger,
``AdS3 Black Holes and a Stringy Exclusion Principle'',
JHEP 9812 005 (1998) [hep-th/9804085];

\bibitem{CHSt} C.G. Callan, J.A. Harvey, A. Strominger,
``World sheet approach to heterotic instantons and solitons'',
Nucl.Phys. B359, 611 (1991); ``Worldbrane actions for string solitons'',
Nucl. Phys. B367 60 (1991);

\bibitem{DGHR} A. Dabholkar, G. Gibbons, J. Harvey,  F. Ruiz-Ruiz,
``Superstrings and solitons'',  Nucl. Phys. B340 33 (1990); 

\bibitem{Ts} Tseytlin, 
``Extreme dyonic black holes in string theory'',
Mod.Phys.Lett. A11 689 (1996) [hep-th/9601177];

\bibitem{GKSg} A. Giveon, D. Kutasov, N. Seiberg, 
``Comments on string theory on $AdS_3$'', Adv.\ Theor.\ Math.\ Phys.\   2, 733 (1998)
[hep-th/9806194];

\bibitem{KSg} D. Kutasov, N. Seiberg, ``More comments on string theory
on $AdS_3$'', JHEP 9904, 008 (1999) [hep-th/9903219];


\bibitem{BORT} J. de Boer, H. Ooguri, H. Robins,
J. Tannenhouser, ``String theory on $AdS_3$'', JHEP 9812 (1998) 026
[hep-th/9812046];

\bibitem{GK} A. Giveon, D. Kutasov, ``Notes on $AdS_3$'',
Nucl. Phys. B621 303 (2002) [hep-th/0106004];

\bibitem{AGS} R. Arguriro, A. Giveon, A. Shomer, ``Superstring
theory on $AdS_3$ and symmetric product'',
JHEP 0012, 003 (2000) [hep-th/0009242];

\bibitem{SW1} N. Seiberg, E. Witten, ``On the D1-D5 system and singular CFT'',
JHEP 9904, 017 (1999) [hep-th/9903224];

\bibitem{Gubser:2002tv}
S.~S.~Gubser, I.~R.~Klebanov and A.~M.~Polyakov,
``A semi-classical limit of the gauge/string correspondence'',
hep-th/0204051;

\bibitem{Frolov:2002av}
S.~Frolov and A.~A.~Tseytlin,
``Semiclassical quantization of rotating superstring in $AdS_5 \times S^5$'',
hep-th/0204226;


\bibitem{KP} E. Kiritsis, B. Pioline, ``Strings in homogeneous
gravitational waves and null holography'', hep-th/0204004;

\bibitem{Penrose} R. Penrose, ``Any spacetime has a plane wave
as a limit'', Differential Geometry and Relativity, {\it  Reidel, Dordrecht, 
1976, pp. 271-275};

\bibitem{BFHP}  M. Blau, J. Figueroa-O'Farrill, C. Hull, G. Papadopoulos,
``Penrose limits and maximal supersymmetry'', hep-th/0201081;

\bibitem{P1} 
J.~Polchinski,
``String Theory. Vol. 1: An Introduction To The Bosonic String'',
{\it  Cambridge, UK: Univ. Pr. (1998) 402 p};

\bibitem{MO} J.~M.~Maldacena and H.~Ooguri,
``Strings in AdS(3) and SL(2,R) WZW model. I: The Spectrum'',
J.\ Math.\ Phys.\   42, 2929 (2001)
[hep-th/0001053];

\bibitem{DFMS} L.J. Dixon, D. Friedan, E.J. Martinec, S.H. Shenker,
``The conformal field theory of orbifolds'', Nucl.Phys. B282, 13 (1987);

\bibitem{KLL} D. Kutasov, F. Larsen, R. Leigh, ``String theory in magnetic monopole
background'', Nucl.\ Phys.\ B550, 183 (1999)
[hep-th/9812027];

\bibitem{LM} F. Larsen, E. Martinec, ``U(1) charges and moduli in the D1-D5 system'',
JHEP 9906, 019 (1999) [hep-th/9905064];

\bibitem{Berkovits:1999im}
N.~Berkovits, C.~Vafa and E.~Witten,
``Conformal field theory of AdS background with Ramond-Ramond flux'',
JHEP 9903, 018 (1999)
[hep-th/9902098];

\bibitem{GKS}  A. Giveon, D. Kutasov, A. Schwimmer,
``Comments on D-branes in $AdS_3$'', Nucl.Phys. B615 (2001) 133
[hep-th/0106005];

\bibitem{PS}  A. Parnachev, D.A. Sahakyan,
``Some remarks on D-branes in $AdS_3$'', JHEP 0110 (2001) 022 [hep-th/0109150];

\bibitem{PST} B. Ponsot, V. Schomerus, J. Teschner, 
``Branes in the Euclidean $AdS_3$'', JHEP 0202 (2002) 016 [hep-th/0112198];

\bibitem{LOP} P. Lee, H. Ooguri, J. Park, 
``Boundary States for $AdS_2$ Branes in $AdS_3$'', hep-th/0112188;

\bibitem{DP} A. Dabholkar, S. Parvizi, ``Dp Branes in PP-wave Background'',
hep-th/0203231;

\bibitem{KNS}  A. Kumar, R.R. Nayak, Sanjay, ``D-Brane Solutions in pp-wave Background'',
hep-th/0204025;

\bibitem{Bak} D. Bak, ``Supersymmetric Branes in PP Wave Background'',
hep-th/0204033;

\bibitem{ST} K. Skenderis, M. Taylor, ``Branes in AdS and pp-wave spacetimes'',
hep-th/0204054.

\bibitem{Billo:2002ff}
M.~Billo and I.~Pesando,
``Boundary states for GS superstrings in an Hpp wave background,''
Phys.\ Lett.\ B {536}, 121 (2002)
[hep-th/0203028].

\end{thebibliography}
\end{document}